\documentclass[a4paper,fleqn,usenatbib]{mnras}

\usepackage{amsmath}	% Advanced maths commands
\usepackage{amssymb}	% Extra maths symbols
\usepackage{graphicx}	% Including figure files
\usepackage{graphicx}
\usepackage[flushleft]{threeparttable}
\usepackage[export]{adjustbox}
\usepackage{upgreek}
\usepackage{times,txfonts}
\usepackage[T1]{fontenc}
\usepackage{ae,aecompl}

\newcommand{\specialcell}[2][c]{%
  \begin{tabular}[#1]{@{}c@{}}#2\end{tabular}}

\title[Is NGC 6535 a Dark Star Cluster Harboring an IMBH?]{MOCCA-SURVEY Database I: Is NGC 6535 a Dark Star Cluster Harboring an IMBH?}

\author[A. Askar et al.]{Abbas Askar$^{1}$\thanks{E-mail: askar@camk.edu.pl},
Paolo Bianchini$^{2}$, 
Ruggero de Vita$^{3}$,
Mirek Giersz$^{1}$,\newauthor
Arkadiusz Hypki$^{4}$
and Sebastian Kamann$^{5}$
\\
$^{1}$Nicolaus Copernicus Astronomical Centre, Polish Academy of Sciences, 
ul. Bartycka 18, 00-716 Warsaw, Poland\\
$^{2}$Max-Planck Institute for Astronomy, Koenigstuhl 17, D-69117 Heidelberg, Germany\\
$^{3}$School of Physics, University of Melbourne, VIC 3010, Australia\\
$^{4}$Leiden Observatory, Leiden University, PO Box 9513, NL-2300 RA Leiden, The Netherlands\\
$^{5}$Institut f{\"u}r Astrophysik, Georg-August-Universit{\"a}t G{\"o}ttingen, Friedrich-Hund-Platz 1, 37077, G{\"o}ttingen, Germany
}

\date{Accepted XXX. Received YYY; in original form ZZZ}

\pubyear{2016}

\begin{document}
\label{firstpage}
\pagerange{\pageref{firstpage}--\pageref{lastpage}}
\maketitle

\begin{abstract} We describe the dynamical evolution of a unique type of dark star cluster model in which the majority of the cluster mass at Hubble time is dominated by an intermediate-mass black hole (IMBH). We analyzed results from about 2000 star cluster models (Survey Database I) simulated using the Monte-Carlo code \textsc{mocca} and identified these dark star cluster models. Taking one of these models, we apply the method of simulating realistic ``mock observations'' by utilizing the \textsc{cocoa} and \textsc{sisco} codes to obtain the photometric and kinematic observational properties of the dark star cluster model at 12 Gyr. We find that the perplexing Galactic globular cluster NGC 6535 closely matches the observational photometric and kinematic properties of the dark star cluster model presented in this paper. Based on our analysis and currently observed properties of NGC 6535, we suggest that this globular cluster could potentially harbour an IMBH. If it exists, the presence of this IMBH can be detected robustly with proposed kinematic observations of NGC 6535.
\end{abstract}


%%%%%%%%%%%%%%%%%%%%%%%%%%%%%%%%%%%%%%%%%%%%%%%%%%

%%%%%%%%%%%%%%%%% BODY OF PAPER %%%%%%%%%%%%%%%%%%

\section{Introduction}\label{sec:int}

Observational evidence has proven the existence of both stellar mass black holes and supermassive black holes. However, no conclusive evidence has been found for the presence of black holes that have masses in the range of $10^{2} \text{--} 10^{5} M_{\odot}$. This elusive mass range for black holes classifies intermediate mass black holes (IMBHs). There are few plausible formation scenarios for IMBHs. They could be remnants of hypothesized metal-free population III stars that formed in the early universe \citep{MadauRees2001} or they may form via dynamical processes in star clusters \citep[e.g.][]{Gierszetal2015,Portegiesetal2004, Gurkan2004, millerhamilton02}. Therefore, the presence of IMBH in globular clusters (GCs) has been the subject of considerable debate \citep[and reference therein]{lutz16, sun13}.

Along with dynamical formation scenarios for IMBH in dense stellar environments, the extrapolation of correlations between supermassive black holes and the masses of their host galaxies 
%along with other properties
to lower mass systems like GCs suggests
that IMBHs could be present in massive clusters \citep{ferrarese, magor,grahamscott}. 

The prevalent methods for detecting IMBH in GCs involve kinematic observations of stars near the cluster centre or searching for electromagnetic radiation from matter accretion onto an IMBH \citep[and references therein]{strader2012}. The presence of an IMBH should have a distinct signature in the line-of-sight velocity dispersion profile from stars close to the IMBH \citep{Gebhardt2000,Gerssen02,Gerssen03,Noyola2008,Lutz2011}, namely a rise in the central velocity dispersion \citep{bahcallwolf}. This method for detecting IMBH is particularly challenging as observing central regions of extremely dense stellar environments like GCs requires instruments with high spatial resolution and accurate velocity measurements \citep{iausbianchini,ruggero16}. Due to the difficulties and uncertainties in measuring the central velocity dispersions of GCs, there have been contradictory results regarding whether or not certain Galactic GCs harbour an IMBH. Certain kinematic observational techniques like integrated light spectroscopy show rising central velocity dispersions for GCs, while resolved stellar kinematics do not confirm this signature associated with the presence of an IMBH (for the specific case of the GC NGC 6388, see \citet{Lutz2011,lutz15} and \citet{Lanzoni2013}; for $\upomega$ Cen see \citet{AndersonMarel2010} and \citet{Noyola2010}). Due to these contradictory results, the detection of IMBH in GCs remains controversial.  

In this paper, we discuss the dynamical evolution of a unique type of star cluster model that emerged from around 2000 models simulated using the  MOCCA code as part of the Survey Database I project \citep{askarGWpaper}. Very early in their dynamical evolution, these clusters form an IMBH via the formation scenario proposed by \citet{Gierszetal2015}. The IMBH mass continues to grow during the cluster evolution and within a Hubble time, the IMBH becomes the dominant mass in the system making up for at least 50\% of the total cluster mass. \citet{BanerjeeKroupa2011} introduced the term ``dark star clusters'' (DSCs) in their work describing the evolution of cluster models with low Galactocentric radius values. In the clusters described by \citet{BanerjeeKroupa2011}, the stars in the outer part would rapidly escape due to tidal stripping leaving a compact cluster held together by a BH subsystem. In the dark star cluster model discussed in this paper, the majority of the mass of the dark star cluster is in the IMBH and is not attributed to the presence of a BH subsystem or a large population of other dark remnants. Recent kinematic observations by \citet{taylor15} of more than a hundred globular clusters in the giant elliptical galaxy NGC 5128 have revealed high mass-to-light ratio values for several of these extragalactic globular clusters. These observational results point towards the presence of a significant dark mass component in these clusters that could potentially be explained by the presence of an IMBH or a subsystem of compact remnants.

In this paper, we provide a detailed and comprehensive description of the observable photometric and kinematic properties of our dark star cluster model, using the strategy of simulating mock observations. This enables us to put forward a direct comparison between models and real data. We find that the observable properties of the dark star cluster model at Hubble time are particularly close to the peculiar Galactic GC NGC 6535 \citep{halford_zaritsky}. We investigate the possibility that NGC 6535 followed an evolution similar to the dark star cluster model described in this paper and we suggest it could be a potential candidate for harbouring an IMBH. The kinematic signature of an IMBH in this relatively small cluster would be easier to detect compared to larger and more massive Galactic GCs and suggested kinematic observations of NGC 6535 may provide clear-cut evidence for the presence of an IMBH in certain GCs. We would like to explicitly state that the dark star cluster model presented in this paper was not specifically tailored to reproduce NGC 6535, however, at 12 Gyrs it has observational properties remarkably similar to NGC 6535. 

In Section \ref{sec:met}, we briefly describe the \textsc{mocca} code for star cluster simulations, the \textsc{cocoa} and \textsc{sisco} codes used to simulate observational data for our simulated cluster model. We provide the details of the evolution of the dark star cluster model that we simulated with \textsc{mocca} and use mock observations to compare the properties of this model with the Galactic GC NGC 6535 in Section \ref{sec:model}. Sections \ref{sec:cocoa} and \ref{sec:sisco} show the results of the simulated observations of our cluster model from \textsc{cocoa} and \textsc{sisco} and we further discuss the possibility of observing an IMBH in NGC 6535 in section \ref{sec:sisco}. In section \ref{sec:last}, we present the main conclusions and discussion of our results.

\section{Method}\label{sec:met}

\subsection{MOCCA}

The \textsc{mocca} (MOnte Carlo Cluster simulAtor) code is a numerical simulation code for long term evolution of spherically symmetric star clusters. \textsc{mocca} is based on H{\'e}non`s Monte Carlo method \citep{Henon1971} that was subsequently improved by Stod\'{o}{\l}kiewicz in the early eighties \citep{Stodolkiewicz1986} and later by Giersz and his collaborators \citep[see][and references therein for a description of \textsc{mocca} and the Monte Carlo Method]{Gierszetal2013}. The Monte Carlo method can be regarded as a statistical way of solving the Fokker-Planck equation in which the cluster is represented as a set of particles. The basic underlying assumptions for the Monte Carlo method are that the cluster needs to be spherically symmetric and that its evolution is driven by two-body relaxation. These assumptions allow the Monte Carlo method to compute the cluster evolution much faster than N-body codes and the implementation of this method allows inclusion of additional physical effects. In addition to relaxation, which drives the dynamical evolution of the system, \textsc{mocca} includes prescriptions for synthetic stellar evolution of all single and binary stars in the system provided by \citet{HurleyPT2000} and \citet{HurleyTP2002} (\textsc{bse} code). \textsc{mocca} also uses direct integration procedures for small $N$ sub-systems using the \textsc{fewbody} code \citep{FCZR2004}. This allows it to accurately model interactions and encounters between different objects. A realistic treatment of escape processes in tidally limited clusters based on \citet{FukushigeHeggie2000} has been implemented in the code. 

\textsc{mocca} has been extensively tested against the results of N-body simulations of star clusters comprising of thousands up to a million stars \citep[][and references therein]{Gierszetal2013,Heggie14,Gierszetal2015,dragon16,Mapelli2016}. The agreement between the results from these two different types of simulations is excellent. This includes the global cluster evolution, mass segregation time scales, treatment of primordial binaries (energy, mass and spatial distributions), and the number of retained neutron stars (NS), BHs and other exotic objects. The output from \textsc{mocca} code is as detailed as direct N-body codes, but \textsc{mocca} is significantly faster than N-body codes and can simulate a realistic GC within a day.

The speed of \textsc{mocca} makes it ideal for simulating a large number of models with different initial conditions. The dark star cluster model presented in this paper was among 1948 models that were simulated as part of the \textsc{mocca} Survey I project \citep{askarGWpaper,Diogo16}. These models cover a wide range of the initial parameter space with different star numbers, metallcities, binary fractions, central densities etc. The details of the initial parameters for the models simulated in MOCCA Survey I database can be found in Table 1 of \citet{askarGWpaper}. The initial conditions for these models were selected arbitrarily to cover a wide range of parameter space and were not selected to reproduce particular cluster models. However, observational properties of the cluster models at Hubble time can reproduce observed properties of most Galactic GCs. A description of the models simulated in the Survey I project can be found in \citet{askarGWpaper}.     

\subsection{COCOA}

To simulate photometric optical observations of our dark star cluster model, we utilize the \textsc{cocoa} (Cluster simulatiOn Comparison with ObservAtions) code \citep{askar2016a}. The \textsc{cocoa} code is being developed to extend numerical simulations of star clusters for direct comparisons with observations. As an input, it uses the snapshots produced by Monte-Carlo or N-body simulations. The snapshots from \textsc{mocca} code contain information about the positions, velocities, stellar parameters and magnitudes of all objects in the star cluster at a specific time during the cluster evolution.
Using the simulation snapshot, \textsc{cocoa} projects numerical data onto the plane of the sky and can create observational data in the form of FITS files. The distance to the cluster, exposure time, resolution and other instrumental specifications for the simulated observational data obtained from \textsc{cocoa} can be adjusted by the user to recreate observations from virtually any telescope \citep{askar2016b}. 

\subsection{SISCO}

To produce a mock kinematic observation of our dark star cluster, we apply the software \textsc{sisco} (Simulating Stellar Cluster Observation) described in \citet{bianchini15}.
The software simulates Integral Field Unit (IFU) observations of stellar systems starting from a dynamical simulation of star clusters and it is optimized to read as an input the projected cluster snapshot derived from the \textsc{cocoa} software.
It assigns to every star a stellar spectrum, based on the stellar parameters, in the waveband of the Ca triplet (8400-8800 \AA) with a resolving power $R\approx20000$ (properties similar to typical IFU instruments like FLAMES@VLT in ARGUS mode).
Then the instrumental setup is customized, including the size of the field-of-view (FOV), the spaxel scale, PSF shape, seeing conditions, signal-to-noise.
The output of the code is a three-dimensional data cube in which each spatial pixel has an assigned spectrum and luminosity information.

\section{The Simulation Model: Dark Star Cluster}\label{sec:model}

\begin{table*}
\centering
\noindent
\caption{Initial parameters of the simulated dark star cluster model.}
\label{table-initial-param}
\noindent
\begin{threeparttable}
%\centering
\begin{tabular}{c|c|c|c|c|c|c|c|c|c}
\hline
N & \specialcell{Mass\\($M_{\odot}$)} & \specialcell{Binary \\Fraction} & Z & IMF & \specialcell{Central Density \\ ($M_{\odot} pc^{-3}$)} & $r_{h}$ (pc) & $r_{t}$ (pc) & $R_{GC}$ (kpc) & Fallback \\ \hline
400000 & $2.42 \times 10^{5}$ & 10\% \tnote{a}           & 0.001           & 2 \tnote{b} & $3.85 \times 10^{6}$ & 1.20                  & 30.0 & 1.94 & On \tnote{c}                \\ \hline

\end{tabular}
\begin{tablenotes}
      \small
      \item[a]The semi-major axis distribution for primordial binaries is uniform in Log(a) with maximum semi-major axis value being 100 AU. Mass ratio ($q$) distribution was uniform for primordial binaries and the eccentricity distribution was thermal.
      \item[b] 2 indicates that the cluster had a two segment initial mass function as given by \citet{Kroupa2001}. The initial masses of the stars were distributed between $0.08 M_{\odot}$ and $100 M_{\odot}$. $r_{h}$ and  $r_{t}$ give the initial intrinsic half-mass radius and tidal radius, respectively. $R_{GC}$ is the Galactocentric radius of the dark star cluster model.
      \item[c] BH kicks are treated with the mass fallback prescription given by \citet{Belczynskietal2002}.
    \end{tablenotes}
    \end{threeparttable}
\end{table*}

From the \textsc{mocca} Survey I simulation models there were 344 models out of 1948 that formed an IMBH (with mass greater than 150$M_{\odot}$) within 12 Gyrs of cluster evolution. From these 344 models, we found 42 models in which the IMBH mass makes up for at least 50\% of the total cluster mass after 12 Gyr of evolution. We categorize these models as dark star clusters harbouring an IMBH. The 42 dark star cluster models emerge from models with diverse initial conditions for parameters such as number of objects ($N=4\times10^{4}$, $1\times10^{5}$, $4\times10^{5}$, $7\times10^{5}$, $1.2\times10^{6}$), metallicity ($Z=0.0002, 0.001, 0.005, 0.006, 0.02$), binary fraction ($0.05, 0.1, 0.3, 0.95$) and initial concentration (initial King concentration parameter $W_{0}=3, 6, 9$). 74\% of the dark star cluster models had high initial concentrations (with $W_{0}=9$) and initial half-mass radius less than 1.5 pc. All 42 dark star cluster models from the survey database have small Galactocentric distance values ranging from 1.1 up to 5 kpc. Most of these dark star clusters models are also characterized by sufficiently small half-mass relaxation times with about 80\% of the cluster models having initial half-mass relaxation times less than 400 Myr. Furthermore, in more than 80\% dark star cluster models, an IMBH of 150$M_{\odot}$ forms within 100 Myr of cluster evolution through the fast scenario described in \citet{Gierszetal2015}. These general trends in the initial properties of the dark star cluster model are easy to understand. Firstly, an IMBH needs to form early enough in the cluster evolution so that its mass can grow significantly, for this low half-mass relaxation times, high central density and concentration are needed. Secondly, a large number of stars need to escape the cluster due to tidal stripping, for this reason the cluster needs to have a small Galactocentric distance value.  Only in 2 dark star cluster models did the IMBH form after 400 Myrs of evolution, in one model the IMBH formed close to 800 Myrs while in the other one the IMBH formed after about 5.8 Gyrs of cluster evolution via the slow scenario explained in \citet{Gierszetal2015}.

\begin{figure}
\vspace*{-0.2cm}
\hspace*{-0.8 cm}
%\begin{center}
%\end{center}
 \includegraphics[height=8.5cm,width=10.6cm]{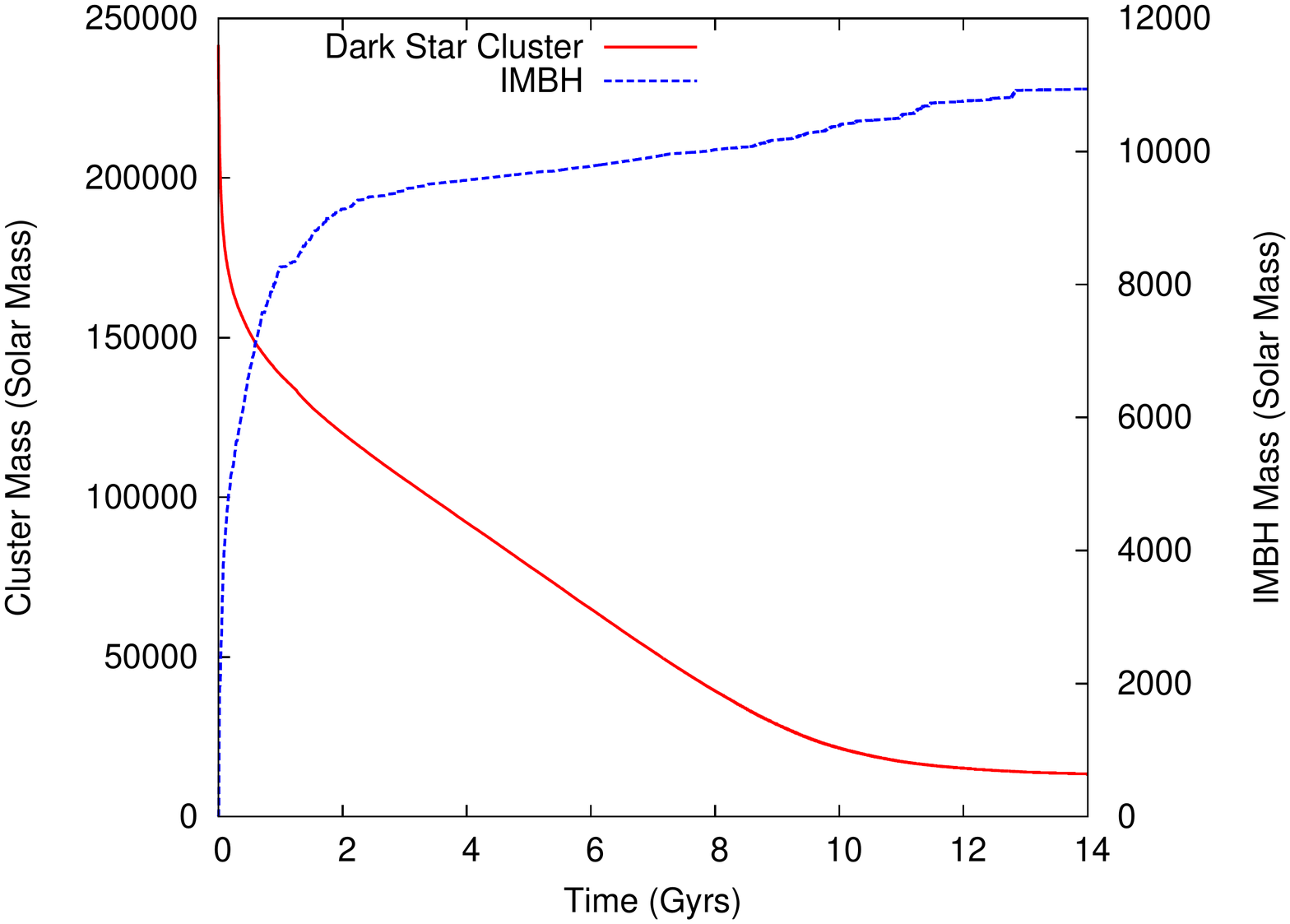} 
 \vspace*{-1.0 cm}
 \caption{The red line shows the evolution of the total mass of the cluster with time. The mass of the cluster decreases almost constantly between 2 and 10 Gyrs of cluster evolution. The blue line shows the build-up of the IMBH mass with time (please refer to the second y-axis for the label for this line). It can be seen that the IMBH mass rapidly increases during the first 2 Gyrs of cluster evolution. }
   \label{fig-time-mass}
%\end{center}
\end{figure}

\begin{figure}
\hspace*{-0.8 cm}
 \includegraphics[height=8cm,width=9.3cm]{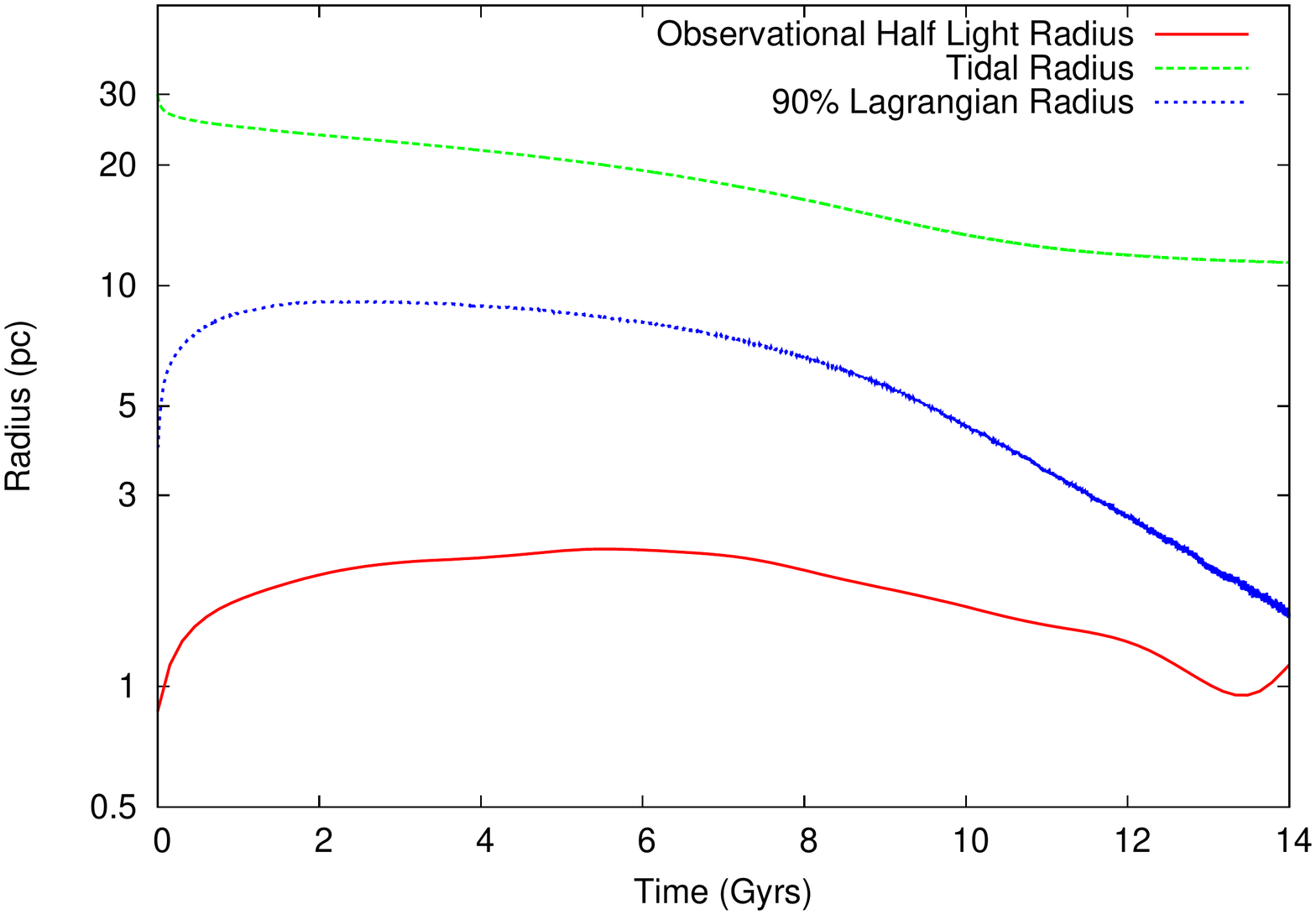} 
\vspace*{-1.0 cm}
 \caption{The figure shows the evolution of the 3D tidal radius, observational 2D half light radius and the 3D 90\% Lagrangian radius of the dark star cluster model with time.}
   \label{fig-time-rh}
\end{figure}

\begin{figure}
\hspace*{-1.0 cm}
 \includegraphics[height=9.0cm,width=10.6cm]{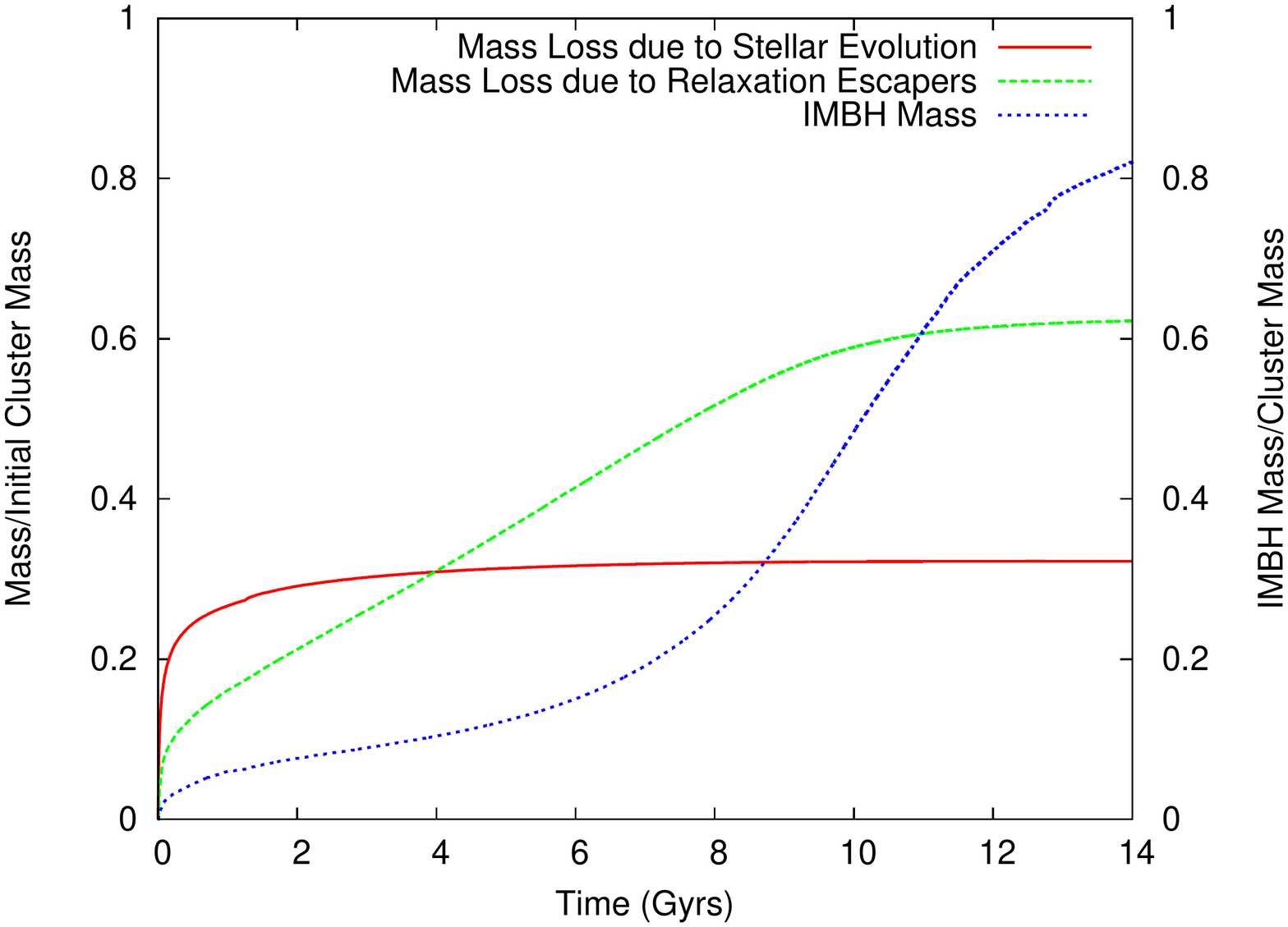} 
 \vspace*{-1.0 cm}
 \caption{The red and green lines in the figure show the ratio between mass loss due to different processes (escapers and stellar evolution) and initial cluster mass. The primary mechanism for mass loss in the dark star cluster model is due to stars escaping because of relaxation after 4 Gyrs of clusters evolution. The blue line shows the ratio between IMBH and cluster mass at that time (please refer to the second y-axis).}
   \label{fig-time-ratio}
\end{figure}

To carry out a more detailed investigation of the evolution of a dark star cluster model and the influence of an IMBH on the 
observable properties of such clusters at 12 Gyrs, we selected a model which had initial conditions representative of most dark star cluster models that emerged in the survey database. The selected model was the most massive model at 12 Gyr in which the the mass of the IMBH at 12 Gyrs made up for at least 70\% of the cluster mass. This particular dark star cluster has initially $N=400,000$ objects with 10\% primordial binary fraction. The initial mass of the cluster model is $2.4 \times 10^{5}$ $M_{\odot}$, with a very high central density of $3.85 \times 10^{6}$ $M_{\odot} pc^{-3}$ and a King concentration parameter value of $W_{0}=9$. The cluster is initially underfilling at a Galactocentric radius of 1.94 kpc. At 12 Gyr, the mass of the cluster is reduced to $1.5 \times 10^{4}$ $M_{\odot}$. The evolution of the cluster mass with time up to 14 Gyrs is shown in Figure \ref{fig-time-mass}. Important initial parameters of the cluster model are given in Table \ref{table-initial-param}. The minimum and maximum initial stellar masses in the simulated model were $0.08 M_{\odot}$ and $100 M_{\odot}$.

\begin{table*}
\centering
\noindent
\caption{Parameters of the simulated dark star cluster model at 12 Gyrs taken directly from the \textsc{mocca} simulation. Values of significant scale radii are also provided in units of arcsec assuming that the dark star cluster is at a distance of 6800 pc from the Sun.}
\label{table-12param}
\noindent
\begin{threeparttable}
%\centering
\begin{tabular}{c|c|c|c|c|c|c|c}
\hline
N & \specialcell{Mass\\($M_{\odot}$)} & \specialcell{Binary \\Fraction} &  \specialcell{Central Density \\ ($M_{\odot} pc^{-3}$)} & \specialcell{$r_{c}$\\pc (arcsec)} & \specialcell{$r_{hl}$\\pc (arcsec)} & \specialcell{$r_{t}$\\pc (arsec)} & \specialcell{IMBH Mass\\($M_{\odot}$)}  \\ \hline
8294 & $1.51 \times 10^{4}$ & 5.8\%            &  $1.30 \times 10^{7}$ & 0.17 (5.16'') & 1.5 (45.5'')                   & 11.92 (361.6'')              & $1.08 \times 10^{4}$             \\ \hline

\end{tabular}

    \end{threeparttable}
\end{table*}

%\vspace{2cm}
The formation of the IMBH in this model is connected with the fast scenario that is explained in \citet{Gierszetal2015}. Here we describe the process of the formation of the IMBH during the early evolution of this particular dark star cluster model. The very dense initial model results in multiple collisions between main sequence (MS) stars \citep{Portegiesetal2004,Mapelli2016} that eventually result in the formation of a massive MS star of 377 $M_{\odot}$. This massive MS star undergoes a direct collision with a BH of 195 $M_{\odot}$ that formed from the evolution of another massive MS star which also underwent mass buildup due to collisions with other stars and mergers in binary interactions. This direct collision between the 377 $M_{\odot}$ MS star and the 195 $M_{\odot}$ BH resulted in the formation of a very massive seed BH in the cluster of 572 $M_{\odot}$ within 9 Myrs of cluster evolution. It is assumed that 100\% of the MS star is accreted onto the BH following a direct collision. This is a rather unrealistic assumption, however, even with less than 50\% of the MS mass being accreted onto a BH, a massive IMBH seed will still be formed \citep{Gierszetal2015}. The mass of the IMBH grows to $1.08 \times 10^{4}$ $M_{\odot}$ at 12 Gyrs, making up for more than 70\% of the cluster mass at this time.

The IMBH continues to gain mass during the evolution of the system, as shown in Figure \ref{fig-time-mass}. The IMBH rapidly gains mass during the first 2 Gyrs of cluster evolution mainly via collisions with other stars and compact objects and mergers after binary-single and binary-binary interactions.
Due to the high initial density, numerous interactions between the stars in the cluster takes place during the first 100 Myrs of cluster evolution which contributes to the very high rate of IMBH mass buildup.

Mass loss due to stellar evolution of massive stars and the formation of an IMBH in this dense system results in the expansion of the half-light radius (Figure \ref{fig-time-rh}) during the first 5 Gyr of cluster evolution (the radius increases from 1.2 pc to 3.65 pc). Due to the small tidal radius of the cluster, a lot of stars escape the cluster due to  tidal stripping and relaxation. After 5 Gyr of evolution, the half-light radius starts to decrease as tidal stripping takes over the expansion connected to energy generated in
the cluster core and the size of the cluster decreases. Escaping stars become the dominant mass loss mechanism for the cluster as the tidal radius of the cluster further decreases. Tidal stripping results in the escape of a large number of objects from the system, with a nearly constant rate until 10 Gyr. At this time, the IMBH mass is about 50\% of the current cluster mass and there are not many stars left in the outer parts of the cluster that would escape easily (see evolution of the 90\% Lagrangian radius in Figure \ref{fig-time-ratio}). The cluster becomes tidally-underfilling again and the rate at which stars are escaping decreases significantly (Figure \ref{fig-time-ratio}). Table \ref{table-12param} reports the relevant properties of the dark star cluster at 12 Gyr.

%\ref{fig-inter} shows the cumulative number of collisions, %binary-single interactions and binary-binary interactions.

%\begin{figure}
%\hspace*{-1.0 cm}
% \includegraphics[height=9cm,width=9.9cm]{time-bh-mass.pdf} 
% \vspace*{-1.0 cm}
% \caption{The figure shows the build-up of the IMBH mass with time. The two lines in the figures show the times when IMBH was a single object or when it was in a binary system with another object.}
%   \label{fig-imbh-mass}
%\end{figure}

%\begin{figure}
%\hspace*{-1.0 cm}
% \includegraphics[height=9cm,width=9.9cm]{time-col-int.pdf} 
% \vspace*{-1.0 cm}
% \caption{The figure shows the cumulative number of collisions, binary-single and binary-binary interactions.}
%   \label{fig-inter}
%\end{figure}

\subsection{Closest Target: NGC 6535}

\begin{figure}
\hspace*{-0.9 cm}
%\begin{center}
 \includegraphics[height=9.4cm,width=10.6cm]{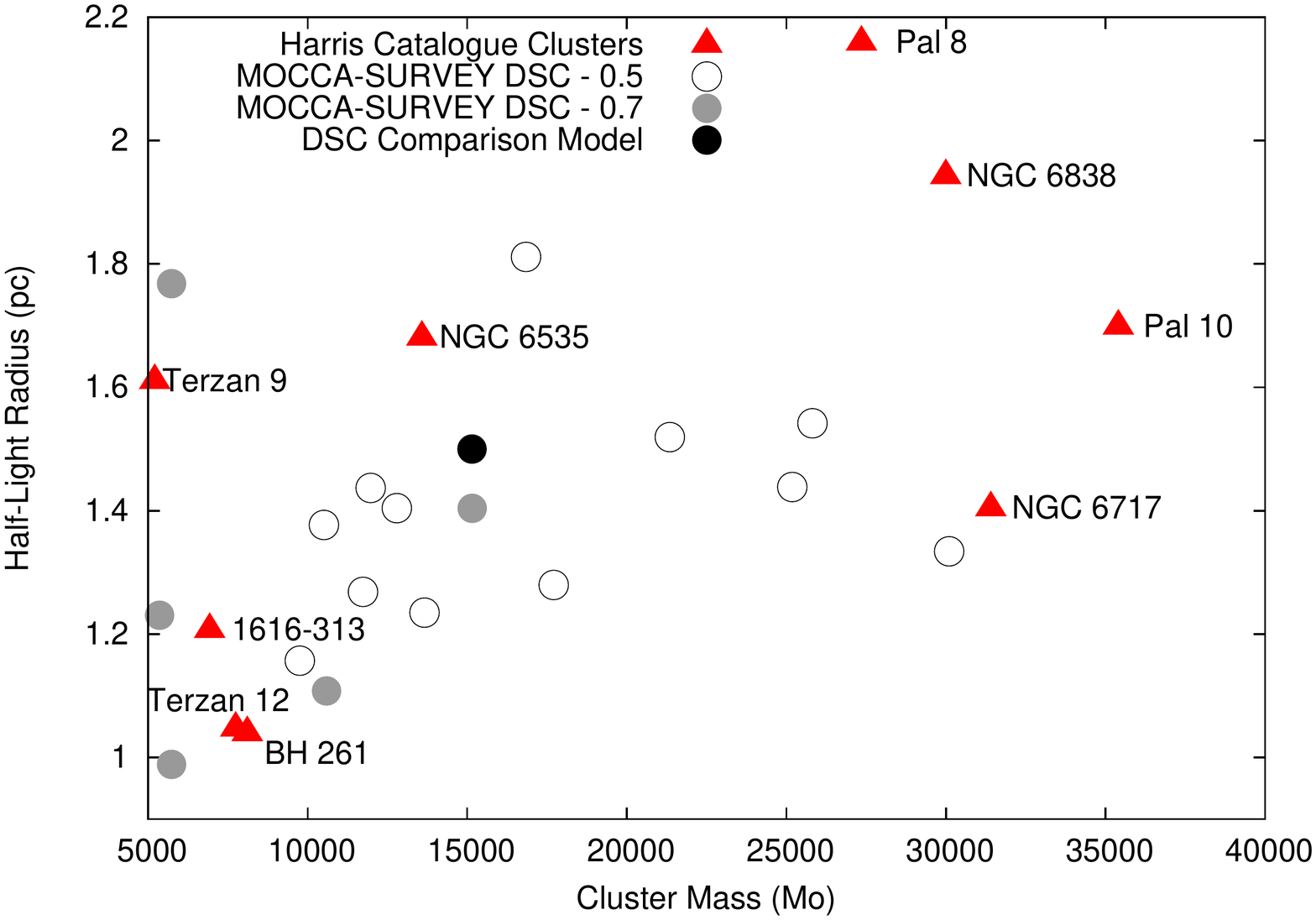} 
 \vspace*{-0.8 cm}
 \caption{Mass vs half-light radius plot for Galactic GCs (red triangles) taken from the \citet[][updated 2010]{Harris1996} catalogue and dark star cluster models in the \textsc{mocca}-Survey database. Open circles show dark star cluster models in which the IMBH mass is larger than 50\% but less than 70\% of the total cluster mass at 12 Gyr. The grey circles show dark star cluster models in which the IMBH makes up for more than 70\% of the cluster mass at 12 Gyrs. The filled black circle represents the dark star cluster model presented in this paper. This model has the closest proximity to NGC 6535 for the models in which the IMBH mass is more than 70\% of the total cluster mass.}
   \label{fig-dark-harris}
%\end{center}
\end{figure}

We compared various properties of our simulated dark star cluster model with the properties of Galactic GCs in the Harris catalogue \citep[][updated 2010]{Harris1996}. Starting from the comparison of the cluster mass and half-light radius, we found that the dark star cluster model simulated with \textsc{mocca} has observational properties at 12 Gyrs that are similar to the Galactic GC NGC 6535. This is a particularly interesting GC. \citet{halford_zaritsky} have described NGC 6535 to be ``sufficiently perplexing'', because of its high dynamical mass-to-light ratio \citep{zaritsky15} and its present day mass function which is ``unusually bottom-light'' \citep{halford_zaritsky}. We would like to stress that the simulated dark star cluster model was not meant to replicate or directly model NGC 6535, but it displays remarkable similarity with the observational properties of this Galactic GC. The comparison between the dark star cluster model and NGC 6535 is summarised below. 

\begin{itemize}
  \item The estimated half light radius for NGC 6535 is 1.68 pc \citep{watkins2015}. The observed half-light radius of our dark star cluster model at 12 Gyrs is 1.5 pc.
%  \vspace{-2 mm}
  \item The core radius of NGC 6535 is 0.71 pc and the concentration parameter $c$ is 1.33, giving a tidal radius of 15.3 pc.  The tidal radius for the dark star cluster model at 12 Gyrs is 11.9 pc.
%  \vspace{-2 mm}
  \item The position of the selected dark star cluster model is in close proximity to NGC 6535 when plotted on a cluster mass vs half-light radius plot of all Galactic GCs. This can be seen in Figure \ref{fig-dark-harris}. For all dark star cluster models in which the IMBH makes up a significant fraction (more than 70\%) of cluster mass at 12 Gyrs, the model presented in this paper has mass and half-mass radii that is closest to NGC 6535. The cluster mass for Galactic GCs was calculated using the absolute visual magnitude ($M_{V}$) and assuming that the $M/L$ is 2 (this is a conservative assumption for NGC 6535).

\begin{figure}
\vspace*{-1.3 cm}
%\begin{center}
\hspace*{-0.8 cm}
 \includegraphics[height=8.8cm,width=9.6cm]{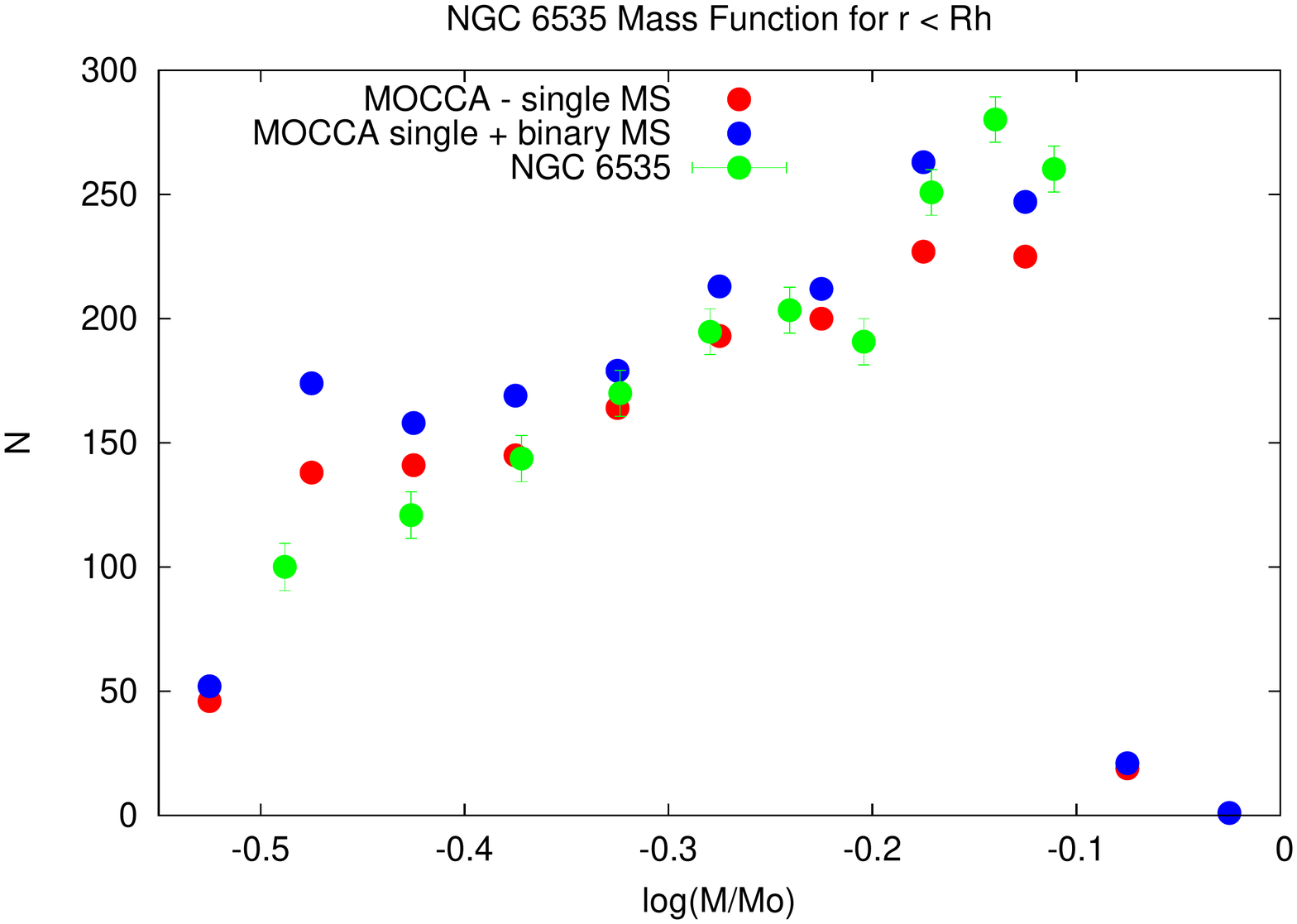} 
 \vspace*{-0.86 cm}
 \caption{The figure above shows the mass function for NGC 6535 obtained by \citet{halford_zaritsky} with green points. The mass function obtained from the dark star cluster model of \textsc{mocca} is shown in red (single stars only) and blue (including binaries) colours . The numbers of stars have been scaled to see the similarity in the shape of the mass function.}
   \label{fig-mf}
%\end{center}
\end{figure}

\item NGC 6535 has a particularly high mass-to-light (M/L) ratio value of about 11 \citep{zaritsky15}. To check whether NGC 6535 may have had a bottom heavy IMF, \citet{halford_zaritsky} used HST data to measure the present day stellar mass function of NGC 6535 and found a dearth of low mass stars in the cluster. We were able to obtain the mass function of our dark star cluster model using the simulation snapshot at 12 Gyr. The shape of the mass function from the dark star cluster model is comparable to the mass function obtained by \citet{halford_zaritsky}, as presented in Figure \ref{fig-mf}. To compare the mass function from the simulation data with the observed mass function and to mimic the observational biases, we took data for single MS stars (plotted in red in Figure \ref{fig-mf}) and we also computed the mass function by including binaries with MS stars and compact object companions and MS-MS binaries which have mass ratio ($q$) less than 0.5 (shown by blue points in Figure \ref{fig-mf}). The high M/L ratio of NGC 6535 and a bottom-light mass function suggest the presence of a significant dark component in this cluster. This dark component could comprise of a large number of compact objects \citep{BanerjeeKroupa2011} like a BH subsystem \citep{dragon16} or it could be an IMBH. Moreover, it has been suggested by \citet{hurley07} that if the ratio between the core and the half-mass radius of a star cluster is larger than 0.5 then ``something out of the ordinary'' may be needed to explain the inner structure of such a star cluster model. NGC 6535 is one of the six globular clusters in our Galaxy in which this ratio is larger than 0.5.
\end{itemize}

The Galactocentric radius of NGC 6535 is 3.9 kpc and it is in close proximity to the Galactic bulge. NGC 6535 has a larger Galacotcentric radius (3.9 kpc) compared to the dark star cluster model (1.94 kpc). However, it is possible that NGC 6535 may have followed a similar evolution to the dark star cluster model despite having a higher Galactocentric radius. This is because the actual Galactic potential is more complicated than the point mass approximation used in \textsc{mocca}. Thus, the actual Galactic potential may have a stronger influence on the dynamical evolution of NGC 6535, comparable to the influence that a point mass Galactic potential would have on a cluster with lower Galactocentric radius and a circular orbit \citep{Madrid2016}. Moreover, if NGC 6535 has an elliptical orbit then the strong influence of the Galactic potential during pericentre passages may also make the evolution of NGC 6535 similar to the dark star cluster model. \citet{halford_zaritsky} remark that it has been suggested by \citet{Paust} that some clusters with close proximity to the Galactic bulge (like NGC 6535) may undergo tidal stripping of low mass stars which could explain the present day mass function of NGC 6535. 

The absolute visual magnitude of the dark star cluster model at 12 Gyr is -4.162 and the observed magnitude of the NGC 6535 is -4.75. The binary fraction for NGC 6535 derived by \citet{milone2012} is  $0.066\pm0.018$ and the binary fraction for the dark star cluster model is 0.058. The observational parameters of NGC 6535 match reasonably well with the parameters derived from the synthetic observations of the dark star cluster model. This suggests that NGC 6535 could be a reasonably good candidate for harbouring an IMBH. Other detailed structural and kinematic similarities between the simulated dark star cluster model and NGC 6535 are further discussed in the subsequent sections.

%\vspace*{-1.0 cm}

\section{Mock Photometry with COCOA}\label{sec:cocoa}
\begin{figure*}
% \vspace*{-2.0 cm}
\begin{center}
 \includegraphics[width=1.0\linewidth]{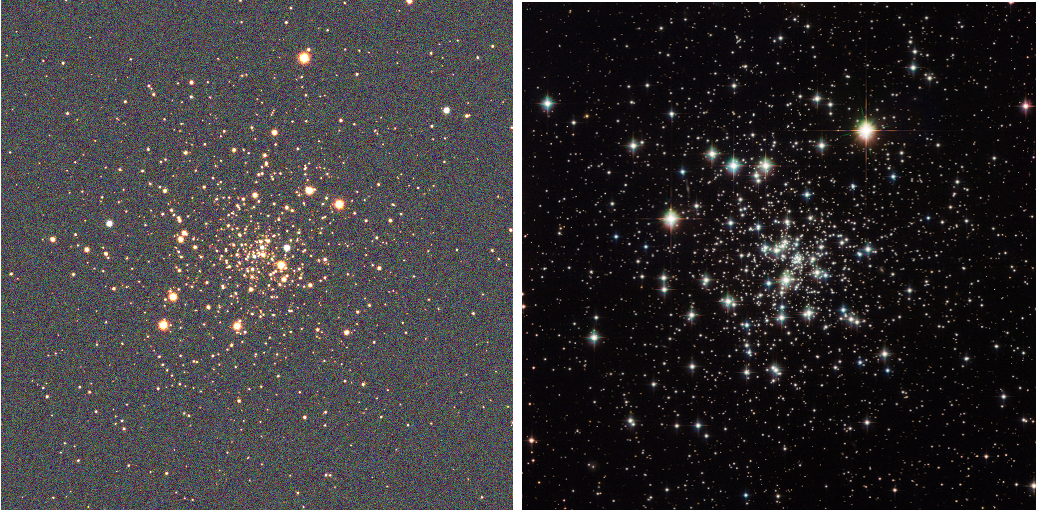} 
% \vspace*{-1.0 cm}
 \caption[]{\textbf{Left Panel:} Mock observation of the dark star cluster model with an 8-meter class telescope (FOV: 2.73 $\times$ 2.73 arcmins). The image was created using \textsc{cocoa} and the 12 Gyr \textsc{mocca} snapshot for the cluster model. The RGB image was obtained by combining synthetic observations in the I,V and B filters. The model was projected at a distance of 6.8 kpc in order to compare it with NGC 6535.
\textbf{Right Panel:} Hubble Image of NGC 6535  from 4 different filters (FOV: 3.31 $\times$ 3.28 arcmins).\protect\footnotemark[1]}
   \label{fig-cluster}
\end{center}

\end{figure*}
\begin{figure}
% \vspace*{-1.0 cm}
\begin{center}
 \includegraphics[width=0.95\linewidth]{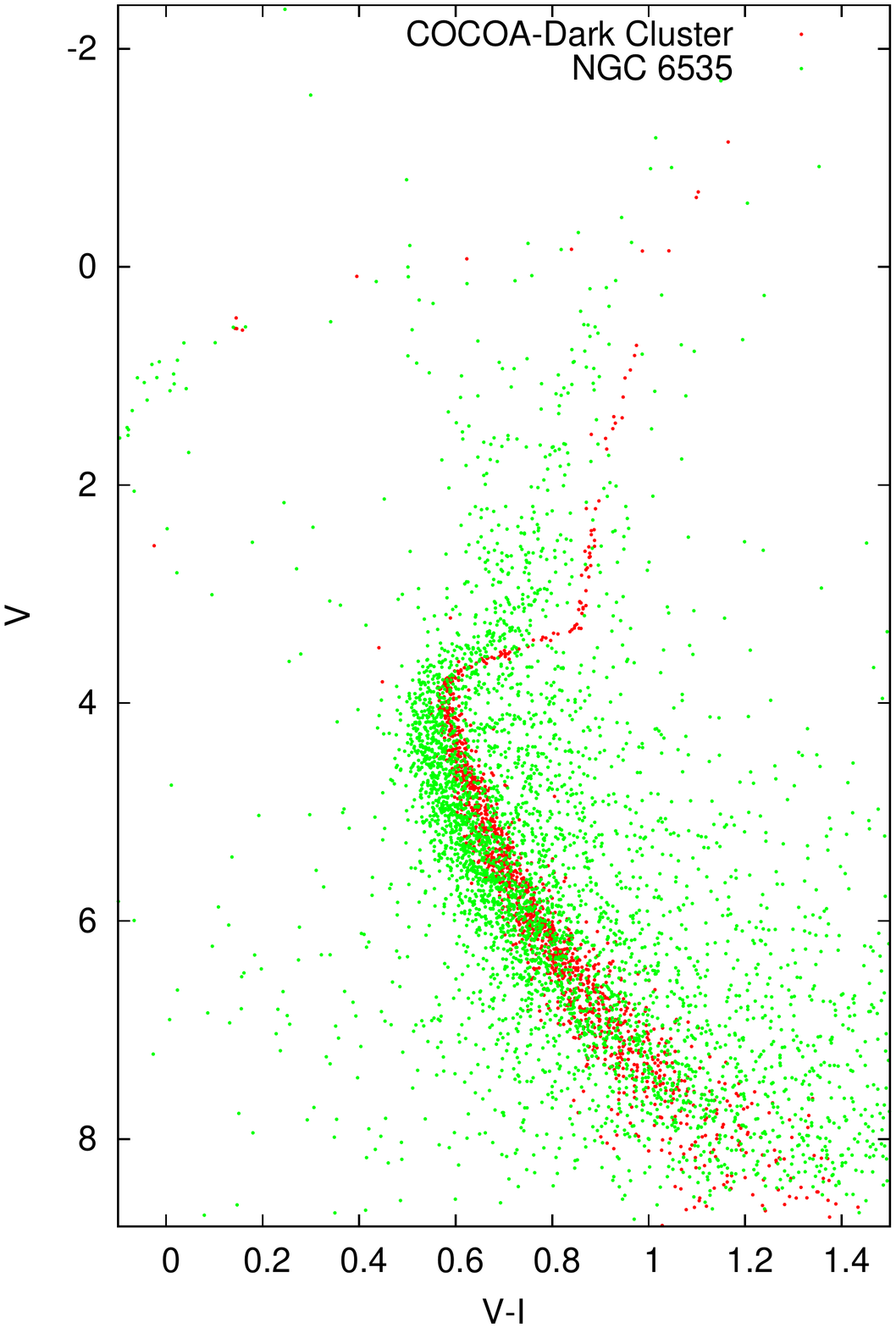} 
 \vspace*{-1.0 cm}
 \caption[]{ V-I vs V colour-magnitude diagram for the dark star cluster model (red points, obtained after simulating observations with \textsc{cocoa}) and NGC 6535 (green points taken from ground-based observations \citep{piotto2002}\protect\footnotemark). The magnitudes are absolute and the observational data has been corrected for extinction. The CMD from NGC 6535 shows that stars are bluer. This can be attributed to the lower metallicity and relatively younger age of NGC 6535.}
   \label{fig-cmd-comparison}
\end{center}
\end{figure}

\large We used the \textsc{cocoa} code in order to simulate idealized synthetic observations of the dark star cluster model and carry out PSF photometry on the simulated images. To compare our dark star cluster model with NGC 6535, we took the simulation snapshot at 12 Gyr and projected it at a distance of 6.8 kpc. To carry out mock photometry of the simulated model, observations of the cluster were simulated with a high spatial resolution optical telescope (8m-class) that had a pixel scale of 0.08 arcsec/pixel. The projected snapshot contained magnitudes for all stars in B,V and I filters; each of these filters was used to simulate separate observations of the dark star cluster model. The left panel in Figure \ref{fig-cluster} shows the composite RGB image of the dark star cluster model that was produced with \textsc{cocoa }using three separate observations in three different filters. The right panel of Figure \ref{fig-cluster} shows an HST image of NGC 6535. It can be seen from the figure that the spatial structure of the dark star cluster model at 12 Gyr is similar to the HST image of NGC 6535. The parameters of the instrument with which the observations are simulated with \textsc{cocoa} are given in Table \ref{table-cocoa}.

\begin{table}
\centering
\caption{Parameters of the instrument for simulated \textsc{cocoa} images. The V-band exposure was 100 times the fraction of direct counts.}
\label{table-cocoa}
\begin{tabular}{|l|l|}
\hline
\textbf{Image Size}  & 2048 $\times$ 2048 pix \\ \hline
\textbf{Distance}    & 6.8 kpc                \\ \hline
\textbf{Pixel Scale} & 0.08 ''/pix            \\ \hline
\textbf{Seeing}      & 0.3''                  \\ \hline
\textbf{PSF}         & Moffat                 \\ \hline
\textbf{GAIN}        & 1.2                    \\ \hline
\textbf{Noise}       & On                     \\ \hline
\end{tabular}
\end{table}

\begin{table*}
%\hspace*{-2.0 cm}
%\centering
\caption{Current observational parameters for the NGC 6535 model and parameters for the dark star cluster (DSC) model at 12 Gyr from \textsc{mocca} and \textsc{cocoa}.}
\label{tab-ngc6535-comp}
\begin{threeparttable}
\begin{tabular}{cccccc}
\multicolumn{1}{l}{}                   & \multicolumn{1}{l}{$\sigma_{0}$ (km/s)  \tnote{a}} & \multicolumn{1}{c}{ c \tnote{b}}   & \multicolumn{1}{l}{Core Radius (pc)} & \multicolumn{1}{l}{Half-light Radius (pc)} & \multicolumn{1}{l}{M/L Ratio} \\ \hline
\multicolumn{1}{|c|}{NGC 6535}         & \multicolumn{1}{c|}{2.40 $\pm$ 0.5 \tnote{c}}                     & \multicolumn{1}{c|}{1.33 $\pm$ 0.16} & \multicolumn{1}{c|}{0.71 (21.6'')}            & \multicolumn{1}{c|}{1.68 (51.0'')}            & \multicolumn{1}{c|}{$11.06^{+2.68}_{-2.12}$ \tnote{d}}             \\ \hline
\multicolumn{1}{|c|}{DSC Model (MOCCA)} & \multicolumn{1}{c|}{37.3}                              & \multicolumn{1}{c|}{1.85}     & \multicolumn{1}{c|}{0.17 (5.16'')}                & \multicolumn{1}{c|}{1.5 (45.5'')}            & \multicolumn{1}{c|}{3.83}           \\ \hline
\multicolumn{1}{|c|}{DSC Model-COCOA King Fit-SBP \tnote{e}} & \multicolumn{1}{c|}{x}                             & \multicolumn{1}{c|}{1.25}     & \multicolumn{1}{c|}{0.52 (16.0'')}            & \multicolumn{1}{c|}{1.24 (37.8'')}                 & \multicolumn{1}{c|}{x}                  \\ \hline

\multicolumn{1}{|c|}{DSC Model-SISCO} & \multicolumn{1}{c|}{15-25 km/s}              & 
\multicolumn{1}{c|}{x}              &
\multicolumn{1}{c|}{x}              &
\multicolumn{1}{c|}{x}              & 
\multicolumn{1}{c|}{17}                  \\ \hline
\end{tabular}
\begin{tablenotes}
       \item[a] $\sigma_{0}$ is the central velocity dispersion in units of km/s.
       \item[b] c defines the concentration parameter given by $\log_{10} \frac{r_{t}}{r_{c}}$, where ${r_{t}}$ is tidal radius and ${r_{c}}$ is the core radius.
       \item[c] Observed value of the central velocity dispersion taken from the Harris catalogue \citep[][updated 2010]{Harris1996}. The observed values for central velocity dispersion are derived from observations of stars further out from the centre of the cluster.
       \item[d]. The observed mass to light ratio for NGC 6535 is taken from \citet{zaritsky15}. 
       \item[e]. Parameters obtained from King model fitting of the surface brightness profile only.
       
\end{tablenotes}
\end{threeparttable}

\end{table*}

\subsection{CMD for the Cluster Model at 12 Gyr}

We carried out PSF photometry on the simulated B,V and I band images to create a catalogue of all stars that could be observed in the cluster model. This was done by employing the automated photometry pipeline in \textsc{cocoa} that utilizes DAOPHOT/ALLSTAR \citep{stetson87,stetson94} to do PSF photometry. By making use of the photometric catalogues from the simulated observation, we were able to obtain the observed colour-magnitude diagrams (CMDs) for our dark star cluster model. We created B-V vs V and V-I vs V CMDs. Figure \ref{fig-cmd-comparison} shows V-I vs V CMD obtained for the dark star cluster model with red points. The instrumental magnitudes obtained from \textsc{cocoa} have been converted to absolute magnitudes. The CMDs show a lack of horizontal branch stars. There are also very few evolved giant stars left in the cluster.

\footnotetext[1]{Link: \mbox{\url{http://www.spacetelescope.org/images/potw1452a/}}\\ Credit: ESA/Hubble \& NASA, Acknowledgement: Gilles Chapdelaine}

The CMD results for the dark star cluster model is compared with observational CMD of NGC 6535 for which the data was obtained through the Rosenberg early ground-based survey \citep{rosen1,rosen2,piotto2002}. Photometric data for NGC 6535 was available in V and I filters. The CMD with colour V-I vs V is plotted from ground based observations of NGC 6535 with green points along with the data from the mock photometry done with \textsc{cocoa} in Figure \ref{fig-cmd-comparison}. The photometric data for NGC 6535 was restricted to stars for which the magnitude error was less than 0.3 and the apparent magnitudes were converted to absolute magnitudes and corrected for reddening. The CMD for NGC 6535 and the dark star cluster model match very well and show how closely the dark star cluster model at 12 Gyr resembles NGC 6535. The V-I colours for NGC 6535 are bluer than the one for the dark star cluster model. The derived metallicity for NGC 6535 is an order of magnitude lower than the metallicity for the simulated dark star cluster model. The CMD obtained from the photometry of the dark star cluster model also closely matches with the CMD from Hubble for NGC 6535 provided in Figure 1 of \citet{halford_zaritsky}. The difference in metallicity and age of NGC 6535 compared to the 12 Gyr snapshot for the dark star cluster model account for the small differences between the two CMDs. In order to check whether decreasing the metallicity would significantly influence the evolution of the dark star cluster model, we simulated the same initial model with a metallicity of $Z=3.24\times10^{-4}$ and found that the cluster follows a very similar evolution to the higher metallicity model. The lower metallicity cluster also goes on to form an IMBH which dominates the cluster mass. The half-light radius and mass function of the lower metallicity cluster model at 10.5 Gyr (estimated age of NGC 6535 taken from \citet{mf-ages}) is comparable to properties of the $Z=0.001$ dark star cluster model at 12 Gyrs.   
\footnotetext{Link: \mbox{\url{http://groups.dfa.unipd.it/ESPG/NGC6535.html}}\\ University of Padova, Stellar Populations Database}

%\begin{figure*}
% \vspace*{-2.0 cm}
%\begin{center}
% \includegraphics[width=0.45\linewidth]{ab-b-v.ps} 
% \includegraphics[width=0.45\linewidth]{ab-v-i.ps} 
% \vspace*{-1.0 cm}
% \caption{\textbf{Left Panel:} B-V vs V colour-magnitude diagram for the dark cluster model obtained using the automated PSF photometry pipeline in COCOA.
%\textbf{Right Panel:} V-I vs V colour-magnitude diagram for the dark cluster model obtained using the automated PSF photometry pipeline in COCOA.}
%   \label{fig-cmd}
%\end{center}
%\end{figure*}

\subsection{Surface Brightness Profile}

\begin{figure}
\hspace*{-1.1 cm}
 \vspace*{-1.2 cm}
 \includegraphics[height=8.2cm,width=9.8cm]{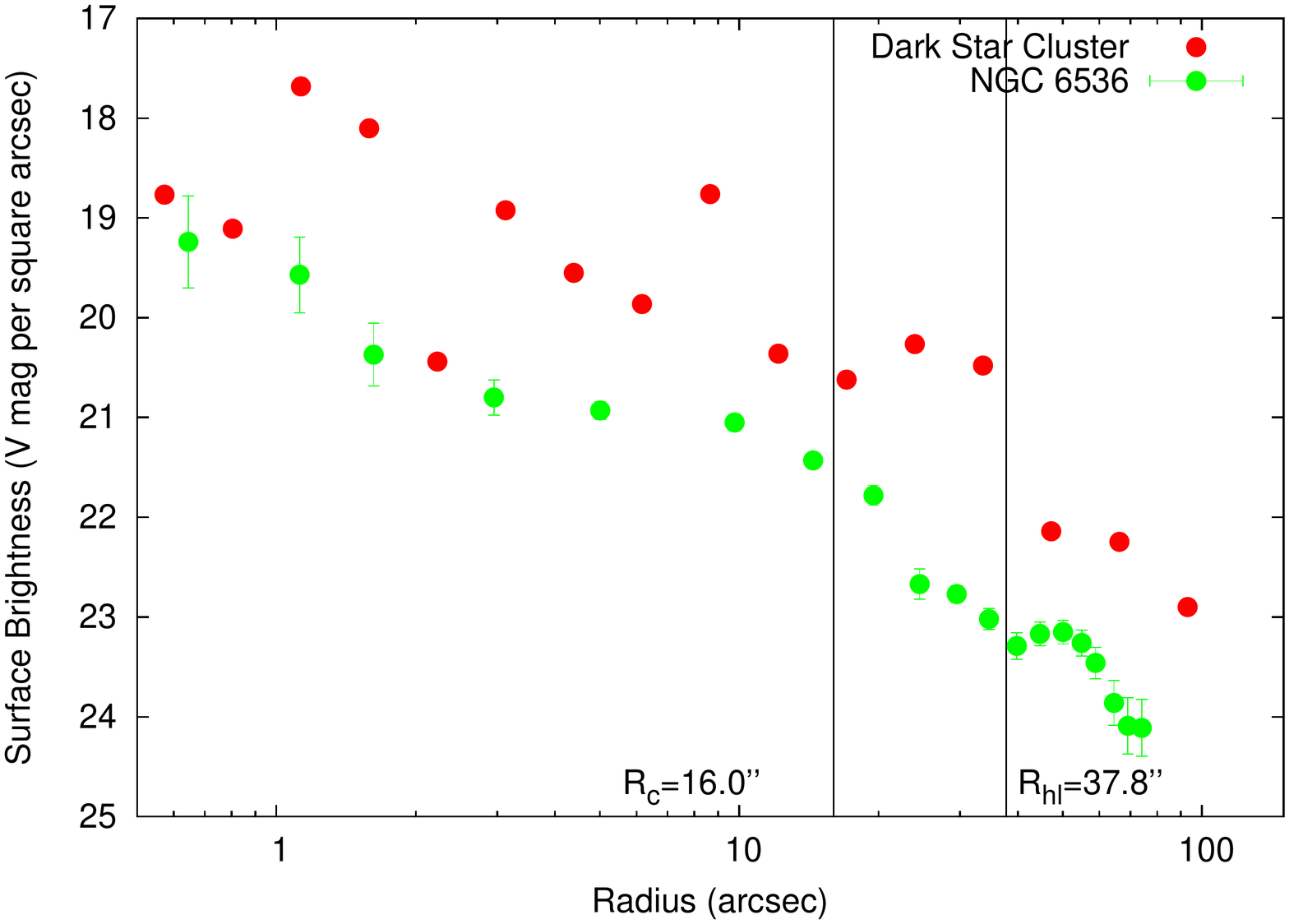} 
 \caption{The red points show the surface brightness profile for the dark star cluster model at 12 Gyr. The y-axis shows the absolute V-band magnitude per square arcsec (for the cluster projected at 6.8 kpc) and the x-axis shows the distance from the centre in arcseconds. The vertical lines in the graph indicate the observed positions of the core and half-light radii. The green points show the surface brightness profile for NGC 6535 obtained by \citet{NG2006}.}
   \label{fig-sbp}
\end{figure}

The surface brightness profile for the dark star cluster model at 12 Gyr is shown in Figure \ref{fig-sbp}. To compare with NGC 6535, the cluster was projected at a distance of 6.8 kpc. The surface brightness profile for our dark star cluster model declines steeply after the radius of 30 arcseconds. The surface brightness profile was fit to the single mass King model to obtain cluster parameters from the projected snapshot. The core radius ($r_{c}$) obtained from the King fitting is 0.53 pc (16 arcsecs). The half light-radius is 1.24 pc (37.8 arcsecs). For comparison, we took the surface brightness profile for NGC 6535 obtained by \citet{NG2006} using HST observations. \citet{NG2006} remark that their observations of NGC 6535 contained very few stars and the measured profile was very noisy. The steepness in the observed surface brightness profile of NGC 6535 is similar to the steepness and shape of the profile of the dark star cluster model with a shift of about 1 mag/$arcsec^{2}$ for points beyond 1''. The core radius and the break radius obtained by \citet{NG2006} from the surface brightness profile gives values of 1.7'' (0.06 pc) and 21.2'' (0.70 pc) which are significantly smaller than more recently observed core and half-light radii for NGC 6535 \citep[][updated 2010]{Harris1996}. 

%\begin{figure}
%\hspace*{-1.2 cm}
% \vspace*{-1.0 cm}
%\begin{center}
% \includegraphics[height=9.0cm,width=11.2cm]{vdp-new2.ps} 
% \vspace*{-1.0 cm}
% \caption{Dispersion profiles for the dark star cluster model at 12 Gyr. The figure shows the line-of-sight velocity dispersion, proper motion dispersion and the dispersion data for NGC 6535 obtained from Watkins et al. 2015. Proper motion dispersion values between 30 and 70 arcsec are comparable for NGC 6535 and the dark star cluster model. }
%   \label{fig-vdp}
%\end{center}
%\end{figure}

%We also computed the line of sight velocity dispersion and proper motion dispersion profiles from our simulation snapshot at 12 Gyrs. To create these profiles, we only selected stars with apparent magnitudes brighter than 20.1 magnitudes, this was done to compare the proper-motion dispersion in our cluster model with that obtained by \citet{watkins2015} (they had a 20.1 limiting magnitude for NGC 6535). The data obtained by \citet{watkins2015} for NGC 6535 was quite limited and did not go close to the center of the cluster. However, dispersion values at outer radii from NGC 6535 data and the dark star cluster model are similar. Figure \ref{fig-vdp} shows the different dispersion profiles. 

We attempted to fit the surface brightness and velocity dispersion profile of the dark star cluster model at 12 Gyr to the single-mass King model \citep{king66} to obtain cluster parameters. Fitting single-mass models has been one of the standard procedures used by observers to obtain cluster parameters. However, due to the presence of an IMBH, fitting both the profiles together was difficult and a reasonable goodness of fit value was not achieved. The presence of a massive compact object or a compact object subsystem in the cluster can result in this difficulty in fitting the \citet{king66} model to the profiles. This problem was also reported and explained by \citet{dragon16} for their clusters which had a presence of a large BH subsystem. In the next section we will describe how the presence of the IMBH in our dark star cluster model significantly influences the kinematic structure, in particular the velocity dispersion profile.
%As a result, the velocity dispersion profile decreases sharply few arcseconds from the center of the cluster (Figure \ref{fig-vdp}). The surface brightness profile is much flatter in the inner part of the cluster (Figure \ref{fig-sbp}) and the core extends to 16 arcseconds. 

%Identifying the proper core radius while fitting both profiles with equal weights does not lead to a good fit. Similar to \citet{dragon16}, we fit the single-mass King model by giving more weight to the velocity dispersion profile and we also fit the model by giving more weight to the velocity dispersion profiles. The cluster parameters obtained from the King fit using the two different weights is provided in Table \ref{tab-ngc6535-comp}. These parameters obtained are not reliable because to improve the goodness of fit value for one profile by giving it more weight resulted in a terrible goodness of fit value for the other profile.  As discussed above, it can be seen from the parameter values that core radius is much larger when more weight is given to the surface brightness profile, also the central velocity dispersion, concentration and M/L ratio values are small. However, giving more weight to the velocity dispersion profile gives a much smaller core radius, central velocity dispersion value is much higher and mass to light ratio is 13.5.  

%\section{Mock Kinematics with sisco}

%=====================================================================
\section{Mock Kinematics with Sisco}\label{sec:sisco}
%\section{Mock kinematics with \texttt{SISCO}}
%=====================================================================

%--------------------------------------------------------------------%
%\begin{figure}
%\begin{center}
% \includegraphics[height=9.0cm,width=11.2cm]{vdp-new3.ps} 
% \caption{Dispersion profiles for the dark star cluster model at 12 Gyr. The figure shows the line-of-sight velocity dispersion and proper motion dispersion of our simulation, overplotted to and the proper motion dispersion profile of NGC 6535 obtained from Watkins et al. 2015. Proper motion dispersion values between 30 and 70 arcsec are comparable for NGC 6535 and the dark star cluster model. }
%   \label{fig-vdp}
%\end{center}
%\end{figure}

\begin{figure}
\hspace*{-0.8 cm}
% \vspace*{-1.2 cm}
\includegraphics[height=8.4cm,width=9.8cm]{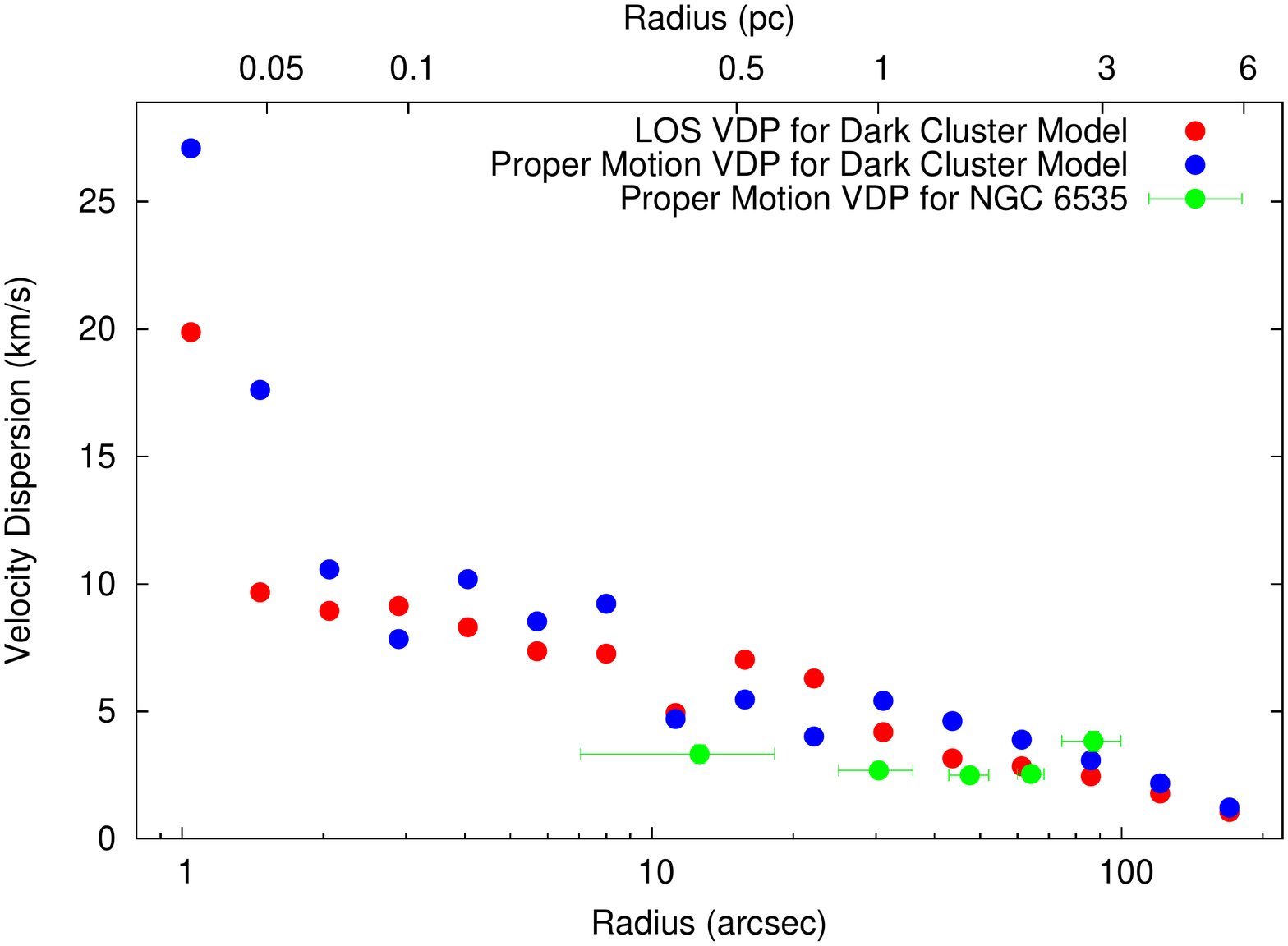} 
 \vspace*{-1.1 cm}
 \caption{The figure shows the comparison between the dispersion profiles for the dark star cluster model at 12 Gyr and available observable dispersion profile for NGC 6535. The red and blue points show the line-of-sight velocity dispersion and proper motion dispersion for the dark star cluster model, green points represent the proper motion dispersion data for NGC 6535 obtained from \citet{watkins2015}.  The dark star cluster model data was restricted to the same limiting magnitude used by \citet{watkins2015} Proper motion dispersion values between 30 and 70 arcsec are comparable for NGC 6535 and the dark star cluster model.} 
 %There is a drop in the VDP for the central bin as only luminous stars were used to compute the VDP and the number of such stars in the IMBH vicinity is very small.
   \label{fig-vdp}
%\end{center}
\end{figure}
%--------------------------------------------------------------------%

Before we explore the details of the kinematics of our simulated cluster, we compare the velocity dispersion profiles (VDP) available in the literature for NGC 6535 with the one obtained from our simulation. The most recent and complete kinematic data of NGC 6535 are the HST proper motions presented in \citet{watkins2015} \citep[see also][]{Bellini2014}. We construct the dispersion profiles (both for the line-of-sight and proper motions) from our simulation considering all the stars brighter than 20.1 magnitudes in the HST F606W filter, in agreement with the procedure followed by \citet{watkins2015}. Figure \ref{fig-vdp} shows the observed dispersion profile and the profile from our simulation. While observational data is missing for the inner 10 arcsec on the cluster, a good agreement is observed at larger radii. Other spectroscopic studies report values of integrated-light velocity dispersion within the half-light radius of $3.27^{+0.67}_{-0.53}$ km s$^{-1}$ \citep{zaritsky15} consistent with the outer values of our profiles and the profile measured by \citet{watkins2015}. Moreover, \citet{watkins2015} find significant radial anisotropy for NGC 6535,  $\sigma_t/\sigma_r\approx0.79$, this is the ratio of the tangential dispersion and radial dispersion on the plane of the sky. Our dark star cluster model is also characterized by a mild radial anisotropy ($\sigma_t/\sigma_r\approx0.97$) when the kinematics is also restricted to the same radial extent and limiting magnitude. However, there is a strong tangential anisotropy in the inner part of the cluster as the average $\sigma_t/\sigma_r\approx1.40$ within about 1.5 arcsec radius from the cluster centre. This tangential anisotropy close to the IMBH can be seen in the higher position of the proper motion dispersion (blue points) compared to the line-of-sight velocity dispersion (red points) for the three innermost radii in Figure \ref{fig-vdp}. This strong tangential anisotropy close to the IMBH is in agreement with theoretical studies on the influence of a central black hole in star clusters and galaxies \citep{young1980,sig95,hb2002,honglee15}.

%\textcolor{red}{QUESTION1:how do you calculate the proper motion dispersion profile? do you use both components on the plane of the sky and then average them (e.g., $\sqrt{\sigma_x^2+\sigma_y^2}$)?}

%\textcolor{red}{QUESTION2: Watkins et al 2015 measure from the proper motion a quite high radial anisotropy for the cluster ($\sigma_t/\sigma_r\approx0.79$, ratio of the tangential dispersion and radial dispersion on the plane of the sky). What is the anisotropy of the simulation?}

Given the missing observational information on the kinematics in the central region, we use our software \textsc{sisco} to simulate integrated-light kinematic measurements. The observational setup used for our mock observation includes a field of view of $20\times20$ arcsec$^{-2}$ (corresponding to $\approx0.4$ $R_h$ around the centre of the cluster), a spaxel scale of 0.25 arcsec/pix, a Moffat PSF with seeing of 1 arcsec and a signal-to-noise ratio of 10 for the spectra (see Table \ref{tab:sisco}). Placing the cluster at a distance of 6.8 kpc, a total of 1484 stars fall within the field of view. From the three-dimensional data cube we extract the velocity dispersion profile, with spatial bins 0.5 arsec wide within the central 2 arcsec and 2.0 arcsec wide for the outer parts.\footnote{The chosen spatial binning is a trade-off between a high enough spatial resolution of the inner area and high signal-to-noise.} For each binned spectra we extract the velocity dispersion (and associated errors) from line-broadening using the penalized pixel-fitting (pPXF) program of \citet{Cappellari2010}. The associated errors and uncertainties for the velocity dispersion measurements are underestimated by the pPXF program. A masking procedure is performed in order to eliminate from the analysis those spaxels which contaminate the kinematics with shot noise  \citep[see][]{bianchini15}. In particular, we discard from the analysis those spaxels for which one bright star contributes for more than 50\% of the total luminosity.

%--------------------------------------------------------------------%
\begin{table}
\begin{center}
\caption{Observational setup of the \textsc{sisco} IFU mock observations}

\begin{tabular}{lc}
\hline\hline
PSF & Moffat\\
seeing & 1 arcsec \\
spaxel scale & 0.25 arcsec/pix \\
field of view & $20\times20$ arcsec$^{-2}$\\
signal-to-noise & 10 \\
\hline

\end{tabular}

\label{tab:sisco}
\end{center}
\end{table}
%------------------------------------------------------------------%

%--------------------------------------------------------------------%
The centre used for our mock observations with SISCO is the centre of coordinates for an inertial systems that is fixed at the beginning of the MOCCA simulation. There are two important radii connected with the influence of a massive black hole in stellar systems, the radius of influence and the wandering radius \citep{baum-a, baum-b,pau2004,stone16}. The radius of influence of the IMBH in the dark star cluster model is about 0.078 pc (2.25 arcsec at an observed distance of 6.8 kpc). The wandering radius which is defined as the offset between the position of the IMBH and the centre of the stellar system for the dark star cluster model is about 0.002 pc or 0.06 arcsec.
The definition of the centre represents one of the main observational challenge for a construction of the velocity dispersion profile and it has been shown that it can significantly influence the results \citep[see e.g.,][]{Noyola2010,Marel2010,bianchini15,ruggero16}.  We therefore explore the effect of picking 10 different locations for the center and construct the corresponding 10 velocity dispersion profiles: one for the real center, three at 1 arcsec distance from the real center, three at 3 arcsec from the real center and three at 5 arcsec from the real center. The three different positions of the FOV centre (($x=0$,$y=R$),($x=R\sqrt{3/2}$,$y=-R/2$) and ($x=-R\sqrt{3/2}$,$y=-R/2$)) for each radial offset $R$ (1,3 and 5 arcsec) are separated at a $120^{\circ}$ on the circle of radius R . This has been done to have different positions of the center for each offset in order to avoid any systematic influence that a few bright stars may cause in the velocity dispersion profile calculation.
%\textcolor{red}{QUESTION: what is the core radius of the cluster from the simulated surface brightness profile (i.e., radius at which the surface brightness is half of the central brightness)? We should quantify the offsets of the center as a fraction of the core radius.}

\begin{figure}
\hspace*{-0.7 cm}
%\begin{center}
% \includegraphics[width=0.4\textwidth]{lum_map.pdf}\quad
 \includegraphics[width=0.5\textwidth]{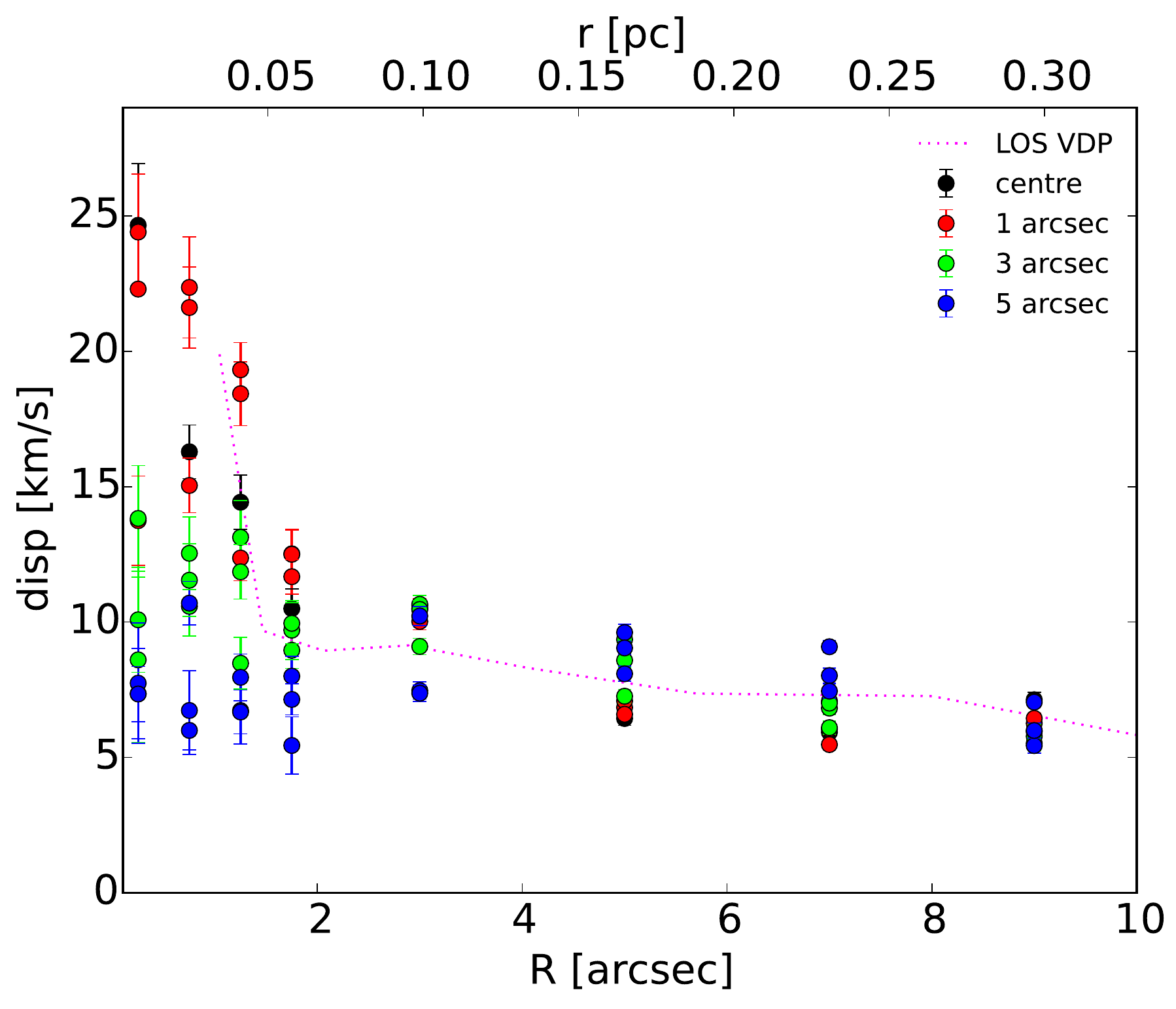}
 \vspace*{-0.35 cm}
 \caption{
% \textbf{Left panel:} Luminosity map obtained for the $20\times20$ arcsec$^{2}$ field of view of our mock IFU \textsc{sisco} observation. Every spaxels correspond to a spectrum from which the kinematic information is obtained from spectral-line broadening. \textbf{Right panel:} 
Integrated-light velocity dispersion profiles obtained from our mock observations. Ten velocity dispersion profiles are shown, indicating the ten different centres used (real centre, three profiles with 1 arcsec offset from the real centre, three profiles with 3 arcsec offsets, and there profiles with 5 arcsec offsets). The black points show the velocity dispersion profile when we use the real centre of the cluster. The dotted magenta line indicates the line-of-sight velocity dispersion profile taken directly from the projected simulation snapshot (red points in Fig. \ref{fig-vdp}). Our kinematic mock observations show that the central peak of the velocity dispersion due to the presence of the IMBH is observable.}
   \label{fig:sisco_kinematics}
%\end{center}
\end{figure}

Fig. \ref{fig:sisco_kinematics} shows the dispersion profiles for the different centers, overlapped to the expected profile constructed directly from the simulation. The profiles constructed using the real center (black points) or a small offset from it (up to 3 arcsec) show a central peak of velocity dispersion, distinctive of the presence of the IMBH, reaching values of $15-25$ km s$^{-1}$. The central rise of the velocity dispersion is erased for a large shift from the real centre (i.e., 5 arcsec), indicating that a careful identification of the centre is needed in order to resolve the central peak. The line-of-sight VDP computed with \textsc{cocoa} from the projected snapshot (Figure \ref{fig-vdp}) and the \textsc{sisco} VDP (Figure \ref{fig:sisco_kinematics}) are in good agreement. In the next section we will use the predicted global velocity dispersion in order to calculate the dynamical mass-to-light ratio and we will show how this can be used as an indicator of the presence of the IMBH, independently from the correct identification of the centre.  

%The data in Figure \ref{fig-vdp} was computed with same limiting magnitude as \citet{watkins2015}, the profile drops steeply at the inner radii. This drop is connected with the absence of luminous stars very close to the IMBH and there is an increased influence of projected stars that come from the outer parts of the cluster. 

\subsection{Virial Estimate of the Dynamical M/L}

%\subsection{Virial estimate of the dynamical M/L}
Using the following virial estimate \citep{Wolf2010}
\begin{equation}
M_{1/2}\simeq930 \left[\frac{\sigma_{los}^2}{km^2\,s^{-2}}\right] \left[\frac{R_h}{pc}\right] M_\odot,
\end{equation}
we can estimate the dynamical mass and dynamical M/L ratio (within the half-light radius) of our simulated cluster based on the kinematic mock observations presented in the previous section. Using a half-light radius of 1.5 pc and an average velocity dispersion of 5 km s$^{-1}$ (lower limit of our mock observations in the central $\approx0.4R_h$ of the cluster\footnote{We caution that the value used for the virial mass estimate should correspond to the one calculated within the half-mass radius.}), we obtain a mass estimate of $M_{1/2}\simeq3.5\times10^4$ $M_\odot$. This dynamical mass estimate is consistent within a factor of two to the total mass of the 12 Gyr snapshot of our simulation (see Table \ref{table-12param}).

Considering a luminosity within the half-light radius of $L_{1/2}=2000$ $L_\odot$, we obtain a $M/L\simeq17$ $M_\odot/L_\odot$. This high dynamical M/L along with a cuspy velocity dispersion profile in the dark star cluster model can be explained as the result of the presence of an IMBH that significantly influences the internal kinematics of the cluster. In order to fully exclude the presence of a BH subsystem in NGC 6535, measurements of the central dispersion profile of NGC 6535 are needed which do not currently exist. 
%Note that \citet{zaritsky15} also report a high value for the mass-to-light ratio of NGC 6535 ($M/L = 11 M_{\odot}/L_{\odot}$).

\subsection{Observing the Possible Kinematic Signature of an IMBH in NGC 6535}

As mentioned earlier, existing searches for signatures of IMBHs using stellar kinematics yield controversial results. The largest discrepancy exists between the results obtained from integrated light measurements and resolved stars. A large part of this controversy can be attributed to our incomplete knowledge of the observed data, especially the PSF. The shape of the PSF determines how strongly bright stars contaminate the integrated light -- as does it determine the fraction of flux contamination from nearby sources in spaxels that seem dominated by individual stars. Unfortunately, at least in ground-based observations, the PSF is not known a priori and has to be modeled from the data. The simulated data presented in the previous section had a known PSF, therefore they necessarily represent a simplification compared to real data. To maximize the reliability of kinematic measurements in the centres of globular clusters, one would ideally combine the results from individual stars and the integrated light. The inclusion of individual stars would also have the added advantage of potentially revealing other kinematical signatures of an IMBH, such as high velocity stars orbiting in the direct vicinity of the black hole, that would be missed by integrated light analyses.

Proper motions as presented in \citet{watkins2015} represent one possibility to obtain results from individual stars. However, as illustrated in Fig.~\ref{fig-vdp}, the existing data do not yet reach far enough into the cluster centre to get meaningful constraints on an IMBH. Alternatively, one may use radial velocities measured from the ground. \citet{Kamann13} presented an approach to extract individual stellar spectra from IFU data via PSF fitting. Not only is this approach capable of suppressing contamination from nearby stars or unresolved light in the extracted spectra, but it also offers several advantages for the analysis of the integrated light \emph{from the same data}. First, it provides an estimate of the shape of the PSF in the data. Second, it can be used to quantify atmospheric effects such as a change of the source positions due to atmospheric refraction. But most importantly, thanks to the knowledge of the PSF, the contributions of the resolved stars can be accurately subtracted from the data, allowing one to study the unresolved light free from their contamination.

We ran the code of \citet{Kamann13} on our \textsc{sisco} simulation of the dark star cluster. The FWHM of the PSF could be recovered to an accuracy of $\sim1\%$. In total about 100 useful stellar spectra were extracted. We measured the velocities of these stars using the IRAF routine \textit{fxcor} and compared them to the output of the simulation. Besides a few outliers due to unresolved binaries the agreement was excellent, highlighting the potential of this approach.

An instrument that would allow the kinematics to be measured from the unresolved light and the resolved stars at the same time needs to combine a high spatial sampling with a high spectral resolution. At the ESO VLT, ARGUS is the only instrument offering this combination. Besides, there are other integral field spectrographs that are not suited for integrated light analyses (because of their relatively low spectral resolution), but would enhance the possibilities for measuring stellar velocities. The AO-assisted IFU SINFONI is one such example \citep{2003SPIE.4841.1548E}, another one is MUSE \citep{2010SPIE.7735E..08B}, the panoramic spectrograph recently installed at the VLT. As shown by \citet{Kamann16}, it allows the simultaneous observation of several thousand stars in the centres of globular clusters. Its upgrade with an AO system is foreseen for next year and will allow one to probe the kinematics around potential black holes in great detail.

%\section{Possible Telescopes/Instruments}

\section{Conclusions \& Discussion}\label{sec:last}

In this paper, we present a new class of star cluster models that emerged from results of thousands of star cluster simulations carried out using the \textsc{mocca} code for star cluster simulations as part of the \textsc{mocca} Survey I project. These particular dark star cluster models form an IMBH via the scenarios explained in \citet{Gierszetal2015} within Hubble time and this IMBH makes up for more than 50\% of the cluster mass at 12 Gyr. Due to the low galactocentric radius values for these cluster models, the low mass stars in the outer parts of the cluster rapidly escape due to tidal stripping. The cluster loses significant mass due to escaping stars and the internal dynamics of the cluster. Moreover, the formation of an IMBH and mass segregation also contribute to this process of sustained mass loss due to escaping stars. The present day mass function for such star clusters is bottom-light.

The structural properties of some of these simulated dark star cluster models (luminosity and radius) show similarities to a sample of low luminosity Galactic GCs. Taking one of these star cluster models, we thoroughly analyzed the photometric and kinematic properties of this model at Hubble time using the strategy of simulating realistic mock observations using the \textsc{cocoa} and \textsc{sisco} codes. This allowed us to compare our dark star cluster model with Galactic GCs and we found that the properties of NGC 6535 are the closest to our dark star cluster model. If our dark star cluster model and NGC 6535 had similar evolutionary histories then it would make NGC 6535 a candidate for harbouring an IMBH. It needs to be stressed that the initial conditions for the models used in the \textsc{mocca}-Survey I were arbitrarily selected and were not tuned to reproduce particular clusters. Despite this, the agreement of the photometric and kinematic properties with NGC 6535 is remarkably good and suggests that this GC may have had a dynamical history similar to our dark star cluster model and consequently it may be harbouring an IMBH. Various observations also suggest that NGC 6535 has a very high mass-to-light ratio value of around 11 \citep{zaritsky15}, which is the highest M/L ratio value for any Galactic GC. Its present day mass function shows that there is a clear absence of low mass stars \citep{halford_zaritsky}, which clearly shows that the high M/L ratio cannot be attributed to a large number of low-mass stars in a scenario where the cluster may have had a bottom-heavy initial mass function. All these observations are in agreement with our dark star cluster model and therefore there maybe a large dark mass component in NGC 6535 in the form of an IMBH.

Future kinematic observations of the central areas of NGC 6535 could unveil a kinematic peak in the velocity dispersion, as predicted by our model. Most observational campaigns that search for kinematic signatures of IMBH in GCs have generally focused on very large and massive globular clusters. There are quite a few problems which can plague the results of such observations for very dense and crowded fields in massive clusters. However, if future campaigns do observe a kinematic peak in a small cluster like NGC 6535 then this would provide clear-cut evidence for the existence of an IMBH. 

\section*{Acknowledgements}

The authors would like to thank the referee for their insightful comments and useful suggestions. AA, MG and AH were partially supported by the Polish National Science Centre (PNSC) through the grant DEC-2012/07/B/ST9/04412. AA would also like to acknowledge support by the PNSC through the grant UMO-2015/17/N/ST9/02573 and partial support from Nicolaus Copernicus Astronomical Centre`s grant for young researchers. PB acknowledges financial support from the International Max-Planck Research School for Astronomy and Cosmic Physics at the University of Heidelberg (IMPRS-HD). RdV acknowledges financial support from the Angelo Della Riccia foundation. SK received funding through BMBF Verbundforschung (project MUSE-AO, grant 05A14MGA).

%%%%%%%%%%%%%%%%%%%%%%%%%%%%%%%%%%%%%%%%%%%%%%%%%%

%%%%%%%%%%%%%%%%%%%%%%%%%%%%%%%%%%%%%%%%%%%%%%%%%%



%\bibliographystyle{mnras}


%%%%%%%%%%%%%%%%%%%%%%%%%%%%%%%%%%%%%%%%%%%%%%%%%%
%%%%%%%%%%%%%%%%%%%%%%%%%%%%%%%%%%%%%%%%%%%%%%%%%%

%%%%%%%%%%%%%%%%% APPENDICES %%%%%%%%%%%%%%%%%%%%%

%\appendix



%%%%%%%%%%%%%%%%%%%%%%%%%%%%%%%%%%%%%%%%%%%%%%%%%%

% Don't change these lines
\bsp	% typesetting comment
\label{lastpage}
\end{document}